\documentstyle[psfig,conf_iap,10pt]{article}
\def\hmpc{\rm \,h^{-1}\,Mpc}

\def\xir{$\xi(r)$\ }
\def\xis{$\xi(s)$\ }
\def\xip{$\xi(r_p,\pi)$\ }
\def\xip{$\xi(r_p,\pi)$\ }
\def\wp{$w_p(r_p)$\ }

\def\kms{\,{\rm km\,s^{-1}}}

\def\ro{r_\circ}
\def\n_med{{\left<n\right>}}

\def\begc{\begin{center} }
\def\endc{\end{center} } 
\def\begf{\begin{figure} }
\def\endf{\end{figure} }

\def\j3{{J_3}}

\begin{document}
\heading{%
%
The Redshift and Real--Space Correlation Functions from the ESP 
Galaxy Redshift Survey
%
} 
\par\medskip\noindent
\author{%

L. Guzzo$^1$, J.G. Bartlett$^2$, A. Cappi$^3$, S. Maurogordato$^4$, 
E. Zucca$^3$, G. Zamorani$^{3,5}$, C. Balkowski$^6$, A. Blanchard$^2$, 
V. Cayatte$^6$, G. Chincarini$^{1,7}$, C.A. Collins$^8$, D. Maccagni$^9$, 
H. MacGillivray$^{10}$, R. Merighi$^3$, M. Mignoli$^3$, D. Proust$^6$, 
M. Ramella$^{11}$, R. Scaramella$^{12}$, G.M. Stirpe$^3$, G. Vettolani$^5$
}
\address{%
Osservatorio Astronomico di Brera, I-23807 Merate, Italy
}
\address{%
Universit\'e L. Pasteur, Obs. Astronomique, F-67000 Strasbourg, France
}
\address{%
Osservatorio Astronomico di Bologna, I-40126 Bologna, Italy
}
\address{Observatoire de Nice, B4229, F-06304 Nice Cedex 4, France
}
\address{Istituto di Radioastronomia del CNR, I-40129 Bologna, Italy
}
\address{Observatoire de Paris, DAEC, F-92195 Meudon, France
}
\address{Universit\`a degli Studi di Milano, I-20133 Milano, Italy
}
\address{Liverpool John--Moores University, Liverpool L3 3AF, UK
}
\address{Istituto di Fisica Cosmica e Tecnologie Relative del CNR, I-20133 Milano, Italy
}
\address{Royal Observatory Edinburgh, Edinburgh EH9 3HJ, UK
}
\address{Osservatorio Astronomico di Trieste, I-34131 Trieste, Italy
}
\address{Osservatorio Astronomico di Roma, I-00040 Monteporzio Catone, Italy
}

\begin{abstract}
We discuss the behaviour of the redshift- and real-space correlation
functions from the ESO Slice Project (ESP) galaxy redshift survey.   
\xis for the whole survey is positive out to $\sim 80\hmpc$, with a
smooth break from a power law. By projecting \xip, we recover the real--space
correlation function \xir, which below $10\hmpc$ is reasonably 
well described by a power law $\xi(r) = (r/r_\circ)^{-\gamma}$, 
with $r_\circ=4.15^{+0.20}_{-0.21} h^{-1}\,$ Mpc and 
$\gamma=1.67^{+0.07}_{-0.09}$.  The same analysis, applied to four
volume-limited subsamples, evidences a small but significant growth 
of clustering with luminosity ($r_\circ$ varies from 3.4 to $5.2 \hmpc$, 
when the luminosity threshold is increased from $-18.5$ to $-20$).  
\end{abstract}
\section{The ESP Survey}
The original goal of the ESO Slice Project (ESP) redshift survey
was essentially twofold.  First, we wanted to produce a measure 
of the galaxy luminosity function in the ``local'' ($z<0.2$) Universe,
with large dynamic range.   Second, we wanted to study the
clustering of galaxies from a survey hopefully not dominated by a single 
major superstructure, as it was the case for the surveys available at the 
beginning of this decade.  
While the ESP luminosity function and its implications are discussed 
in detail in \cite{Zuc97} and summarised also in these proceedings by 
Elena Zucca, here we shall summarise our results on the two--point 
correlation properties of galaxies.

The ESP survey covers a strip of sky $1^\circ$ 
thick (DEC) by $\sim30^\circ$ long (RA) (with an unobserved $5^\circ$ sector
inside this strip), in the SGP region.  The target galaxies were selected 
from the EDSGC \cite{Hey}, and the final catalogue contains 3342 redshifts,
corresponding to a completeness of 85\% at a magnitude limit $b_J=19.4$
\cite{Vet98}.  We use $H_o=100\kms$ Mpc$^{-1}$ and $q_o=0.5$.
Magnitudes are K--corrected as described in \cite{Zuc97}.

\section{Clustering in Redshift Space}
\begin{figure}
\centerline{\vbox{
\psfig{figure=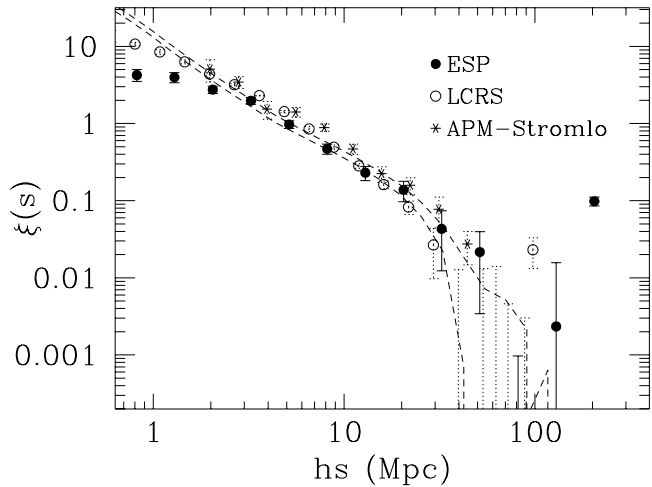,height=7.cm,angle=270}
}}
\caption[]{The redshift-space correlation function from the whole ESP
magnitude--limited catalogue, compared to previous data (see text).
}
\label{xi_mlim_surv}
\end{figure}
The filled circles in Figure ~\ref{xi_mlim_surv} show \xis computed from
the whole ESP magnitude--limited sample using the
J3--weighting technique, (e.g. \cite{F94a}).  Note the smooth decay
from the power law at large separations, with correlations going to
zero only around $80\hmpc$, and perhaps some evidence for a positive 
fluctuation around
$\sim 200\hmpc$.  Below $\sim 3\hmpc$, \xis tends to flatten. 
This is the effect of  
redshift--space damping from pairs within virialized structures 
and interestingly it is less prominent when \xis is
computed for volume--limited subsamples \cite{Guz98a}.
In the same figure, we show \xis 
from the Las Campanas (LCRS, \cite{Schect}), and APM--Stromlo \cite{Lov}
redshift surveys.  There is a rather good agreement of the three independent
data sets between 2 and $20\hmpc$ (where $\xi(s)=(s/s_\circ)^{-\gamma}$, with
$s_\circ\sim 6 \hmpc$ and $\gamma\sim-1.5$), with perhaps a hint for more power
on larger scales in the 
blue--selected ESP and APM--Stromlo.  In addition, the dashed lines describe
the real--space \xir obtained by de-projecting the angular correlation
function $w(\theta)$ of the APM Galaxy Catalogue \cite{Baugh}, for two 
different assumptions on the evolution of clustering.  It is rather 
interesting to note the degree of unanimity between the angular and 
redshift data in indicating both the large--scale shape and the 
zero-crossing scale ($40-90\hmpc$) of galaxy correlations.  
The small amplitude difference between the redshift and 
real--space correlations on scales $>5\hmpc$ is also remarkable, because
it suggests a small amplification of \xis due to streaming flows, and thus
a low value of $\beta=\Omega_o/b$.  

\section{Clustering in Real Space}
\begin{figure}
\centerline{\vbox{
\psfig{figure=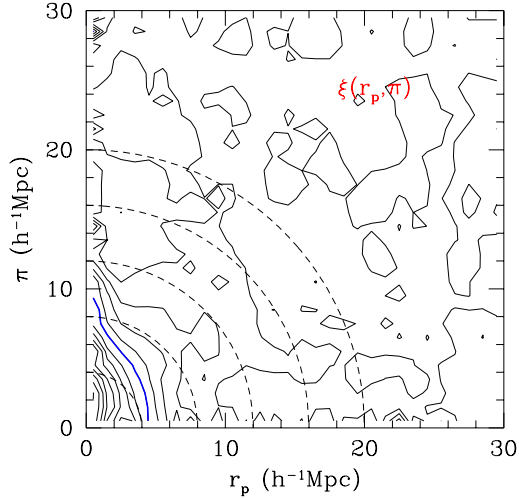,height=7.cm}
}}
\caption[]{\xip contour map for the whole ESP survey. 
The heavy contour corresponds to $\xi=1$.
}
\label{csipz}
\end{figure}

Redshift distortions produced by peculiar velocities, are usually studied 
through the function \xip \cite{Guz97}, 
that describes galaxy correlations as a function
of two variables, one perpendicular ($r_p$), and the other parallel ($\pi$),
to a sort of mean line of sight defined for each galaxy pair.  
In Figure~\ref{csipz}, we show \xip computed for the whole ESP survey, using
the same technique used for \xis.  From this figure, one can clearly see the
small--scale stretching of the contours along $\pi$, produced by the relative
velocity dispersion of pairs within structures.
Projecting \xip onto the $r_p$ axis, one gets the function $w_p(r_p)$,
which is independent from the redshift--distortion field,
and is analytically integrable for a power--law 
$\xi(r)= (r/r_\circ)^{-\gamma}$.  In this case, the values of 
$r_\circ$ and $\gamma$ that best reproduce the data can be evaluated
through an appropriate best--fitting procedure.
By applying this to the
map of Figure~\ref{csipz}, we recover 
$r_o=4.15^{+0.20}_{-0.21} h^{-1}\,$ Mpc and $\gamma=1.67^{+0.07}_{-0.09}$,
a value that is slightly smaller than that measured from the red-selected
LCRS \cite{Lin_th}.
\begin{figure}
\centerline{\vbox{
\psfig{figure=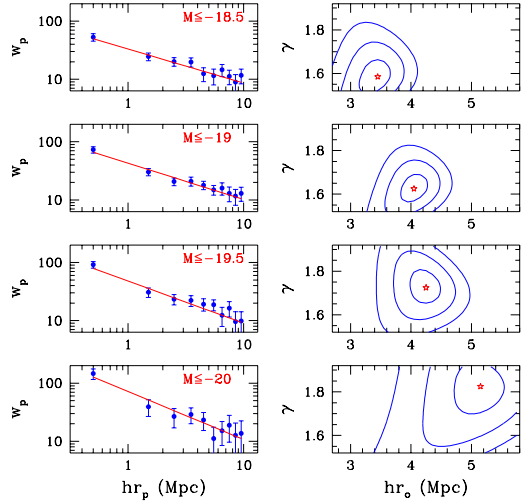,height=7.cm}
}}
\caption[]{Fits of a power--law \xir to \wp for four volume--limited 
samples.
}
\label{wpfit_vlim}
\end{figure}
Finally, we have investigated the stability of \xir as a function of
the sample mean luminosity, by computing \wp for four volume--limited
subsamples.  Figure \ref{wpfit_vlim} shows the result of the power--law
fit to \wp, evidencing a weak, but significant growth of $\ro$ and $\gamma$
when more and more luminous galaxies are selected.  Interestingly, 
this behaviour
is completely masked if one looks at \xis \cite{Guz98a}, the reason being
the rather different small--scale distortion acting in the four cases, that
basically compensates the (real--space) growth of clustering.  
This shows the importance of separating out dynamical
effects when comparing the clustering of different classes of objects
through the two--point correlation function.
\vfill

\begin{iapbib}{99}{
\bibitem{Baugh} Baugh, C.M., 1996, \mn 280, 267
\bibitem{F94a} Fisher, K.B., Davis, M., Strauss, M.A., 
   Yahil, A., \& Huchra, J.P. 1994a, MNRAS, 266, 50 (F94a)
\bibitem{Hey} Heydon--Dumbleton, N.H., Collins, C.A., MacGillivray, H.T., 1989,
           \mn 238, 379
\bibitem{Guz97} Guzzo, L., Strauss, M.A., Fisher, K.B., Giovanelli, R., 
and Haynes, M.P. 1997, \apj, 489, 37
\bibitem{Guz98a} Guzzo, L., et al. (The ESP Team), 1998a, \aeta, submitted
\bibitem{Lin_th} Lin, H., 1995, PhD Thesis, Harvard University
\bibitem{Lov} Loveday, J., Peterson, B. A. Maddox, S. J., Efstathiou, G., 
1996, \apj Suppl. 107, 201
\bibitem{Schect} Schectman, S.A., Landy, S.D., Oemler, et al., 1996, \apj 470,
 172
\bibitem{Vet97} Vettolani, G., et al. (the ESP Team), 1997, \aeta 325, 954
\bibitem{Vet98} Vettolani, G., et al. (the ESP Team), 1998, \aeta Suppl., 130, 323
\bibitem{Zuc97} Zucca, E., et al. (the ESP Team), 1997, \aeta, 326, 477
}
\end{iapbib}
\end{document}